\documentclass[aps,prb,reprint,twocolumn,amsmath,amssymb,citeautoscript,longbibliography]{revtex4-2}
\usepackage{graphicx}
\usepackage{dcolumn}
\usepackage{bm}
\usepackage{latexsym,epsfig}
\usepackage{graphicx}
\usepackage{verbatim}
\usepackage{comment}
\usepackage{amsmath}
\usepackage{amssymb}
\usepackage{physics}
\usepackage{stmaryrd}
\usepackage{color}
\usepackage{graphicx}
\usepackage{epstopdf}
\usepackage{grffile}
\usepackage{lipsum}
\usepackage{enumitem}
\usepackage{ulem}
\usepackage[dvipsnames]{xcolor}
\DeclareGraphicsExtensions{.eps}

\newcommand{\beq}{\begin{equation}}
\newcommand{\eeq}{\end{equation}}
\newcommand{\bea}{\begin{eqnarray}}
\newcommand{\eea}{\end{eqnarray}}
\newcommand{\ben}{\begin{eqnarray*}}
\newcommand{\een}{\end{eqnarray*}}
\newcommand{\bfig}{\begin{figure}}
\newcommand{\efig}{\end{figure}}

\usepackage{hyperref}
\hypersetup{
    colorlinks=true,      
    urlcolor=blue,
    citecolor=blue,
    linkcolor=blue
}

\usepackage{booktabs}
\usepackage{array}
\usepackage{lipsum} 
\usepackage{array} 

\begin{document}
\title{Quasiperiodicity-Engineered Re-entrant Localization-Delocalization aspects in a Diamond Lattice} 








\author{Ranjini Bhattacharya}
\email{ranjinibhattacharya@gmail.com}
\affiliation{Institute of Physics, Sachivalaya Marg, Bhubaneswar-751005, India}
\affiliation{Homi Bhabha National Institute, Training School Complex, Anushaktinagar, Mumbai 400094, India}

\author{Souvik Roy}
\email{souvikroy138@gmail.com}
\affiliation{School of Physical Sciences, National Institute of Science Education and Research, Jatni 752050, India}
\affiliation{Homi Bhabha National Institute, Training School Complex, Anushaktinagar, Mumbai 400094, India}

\date{\today}

\begin{abstract}
We investigate localization in a quasiperiodically engineered diamond lattice with strand-dependent Aubry--André--Harper onsite modulations, highlighting the decisive roles of the modulation ratio $s$ and the averaged potential on the middle strand. The upper strand hosts the primary potential $\lambda$, the lower strand carries a weaker modulation $\lambda/s$, and the middle strand follows their average, generating a correlated quasiperiodic landscape across each plaquette. By tuning $\lambda$ for selected values of $s$, we probe spectral and eigenstate properties via the inverse participation ratio (IPR), normalized participation ratio (NPR), and fractal dimension $D_2$. We uncover a pronounced re-entrant localization behavior, where eigenstates repeatedly switch between extended and localized regimes, which persists only within a finite range of $s$ and crucially relies on the averaged potential construction. This unconventional sequence arises from the interplay of $s$, the correlated potential, and the intrinsic diamond geometry, producing a highly nontrivial interference landscape. Our results reveal localization physics beyond the standard Aubry--André paradigm, further supported by the evolution of extended states, system-size scaling of $\langle \mathrm{NPR} \rangle$ and $\langle D_2 \rangle$, and dynamical signatures from the time-dependent root-mean-square displacement, confirming the robustness of the re-entrant transitions.
\end{abstract}

\maketitle

\section{Introduction}


Anderson localization is one of the most fundamental phenomena in condensed matter physics, describing the suppression of wave transport in disordered systems due to quantum interference effects. First proposed by P. W. Anderson in 1958 while analyzing spin-resonance experiments in doped silicon, the concept explains how random disorder can strongly reduce quantum tunneling and confine particles such as electrons and photons within a lattice~\cite{ander1, ander2}. In contrast to periodic systems, where single-particle states form extended Bloch waves, disorder can induce exponential localization of wave functions through multiple scattering and destructive interference~\cite{ander3, ander4, ander5}. A notable consequence is that in one- and two-dimensional systems with uncorrelated random disorder, even arbitrarily weak disorder localizes all eigenstates, preventing localization–delocalization transitions and eliminating the possibility of mobility edges separating extended and localized states. In three dimensions, however, sufficiently weak disorder can lead to mobility edges, allowing the coexistence of localized and extended states~\cite{c1, c2, c3, c4}. The concept of mobility edges is important because of its potential applications in electronic transport, switching devices, and thermoelectric systems~\cite{c5, c6,c7}.

Quasiperiodic lattices provide an intermediate platform between perfectly ordered and fully disordered systems, where localization transitions can occur even in low dimensions~\cite{c8}. Owing to their experimental accessibility, quasiperiodic structures have been realized in platforms such as optical lattices, photonic lattices, optical cavities, and superconducting circuits, enabling the observation of several intriguing phenomena including Anderson localization~\cite{c9, c10, c11}, Bose glass phases~\cite{c12, c13, c14}, long-range order, and many-body localization~\cite{c15, c16, c17, c18, c19}.

Another important platform for studying Anderson localization is provided by quasiperiodic systems, which lie between perfectly periodic crystals and completely random disordered lattices. Unlike random systems, quasiperiodic structures possess deterministic spatial correlations while lacking translational symmetry, leading to physical properties absent in both periodic and disordered systems. These systems exhibit phenomena such as localization–delocalization transitions in low dimensions, critical eigenstates, and fractal energy spectra. Among various models, the Aubry–André (AA) or Aubry–André–Harper (AAH)~\cite{aubry1, aubry2} model has emerged as a paradigmatic framework for investigating localization in quasiperiodic systems. In its standard one-dimensional form, the model introduces an incommensurate modulation in the on-site potential and exhibits a sharp metal–insulator transition~\cite{mi1} at a finite critical quasiperiodic strength due to its self-dual symmetry. Below this critical point all eigenstates remain extended, while above it they become exponentially localized, implying the absence of an intermediate phase or energy-dependent mobility edge. However, several generalizations of the AAH model, such as longer-range hopping, modified quasiperiodic potentials, or coupled chains, can break this self-duality and lead to intermediate mixed phases where localized and extended states coexist in the spectrum separated by mobility edges~\cite{mb1, mb2, mb3, mb4, mb5, mb6, mb7, mb8}. These quasiperiodic lattices therefore provide an ideal platform for exploring multifractal eigenstates, critical spectra, and mobility-edge phenomena.

Contrary to the conventional understanding of localization transitions, recent studies have revealed the intriguing phenomenon of reentrant localization (RL)~\cite{rel1, rel2, rel3, rel4} in quasiperiodic systems. Traditionally, once a system undergoes a localization transition, a further increase in disorder or quasiperiodic strength is expected to enhance localization. However, several works have shown that certain eigenstates may reenter an extended phase after initially becoming localized. Early evidence of this behavior was reported by Hiramoto and Kohmoto~\cite{H1}, where states near the band edges exhibited extended–localized–extended transitions as the modulation strength increased. More recently, reentrant localization has been observed in the interpolating Aubry–André–Fibonacci (IAAF)~\cite{rel4, aaf1, fibo1, fibo2, fibo3, fibo4} model and related quasiperiodic lattices. In these systems, mechanisms such as hopping dimerization, staggered quasiperiodic potentials, or modified correlations can drive successive localization transitions, resulting in intermediate phases containing mobility edges. In such cases, the system may first undergo a transition where all eigenstates become localized; with further increase in modulation strength, some states delocalize again before eventually returning to a fully localized phase at larger disorder strengths. This unusual behavior leads to multiple critical points and mobility edges across the spectrum and highlights the rich localization physics that can arise in quasiperiodic and correlated systems.

In this work, we investigate localization in a quasiperiodically modulated diamond lattice by tuning the quasiperiodic strength $\lambda$, with particular emphasis on the modulation ratio $s$ and the averaged onsite potential. The upper and lower strands host AAH-type potentials of strengths $\lambda$ and $\lambda/s$, respectively, while the middle strand follows their average, generating a correlated quasiperiodic landscape. The spectral and eigenstate properties are characterized using the inverse participation ratio (IPR), normalized participation ratio (NPR), and fractal dimension $D_2$. We uncover pronounced re-entrant localization–delocalization (LD) transition, where eigenstates repeatedly alternate between extended and localized regimes with increasing $\lambda$. This behavior persists only within a finite range of $s$, underscoring the key roles of the tunable parameter $s$ and the averaged potential on the middle strand in stabilizing the re-entrant phase. The effect is significantly enhanced when averaging is restricted to a narrow energy window $[-0.1,0.1]$, compared to a broader range such as $[-0.5,0.5]$. Finite-size analysis further confirms the robustness of these features.

We further analyze the evolution of extended states with $\lambda$, along with the system-size scaling of  $\langle \mathrm{NPR} \rangle$ and $\langle D_2 \rangle$, and investigate dynamical signatures via the time-dependent root-mean-square displacement~\cite{dyna1,dyna2,dyna3,dyna4,dyna5}, all consistently supporting the re-entrant nature of the transition. Importantly, in contrast to conventional mechanisms involving both onsite and hopping modulations, the re-entrant LD transition here arises purely from strand-dependent onsite quasiperiodic potentials. The interplay between the modulation ratio $s$ and the correlated averaged potential is thus identified as the key mechanism driving the observed unconventional localization behavior.

The paper is organized as follows. In Sec.~I, we introduce the motivation and objectives of the study. Sec.~II presents the quasiperiodic diamond lattice model and the methodology used to characterize localization--delocalization transitions. The main results are discussed in Sec.~III. Finally, Sec.~IV summarizes the key findings and their implications, with additional details provided in the Appendix.

\section{Theoretical Formulation}
\begin{figure}[t]
    \centering
   \includegraphics[width=1.0\linewidth]{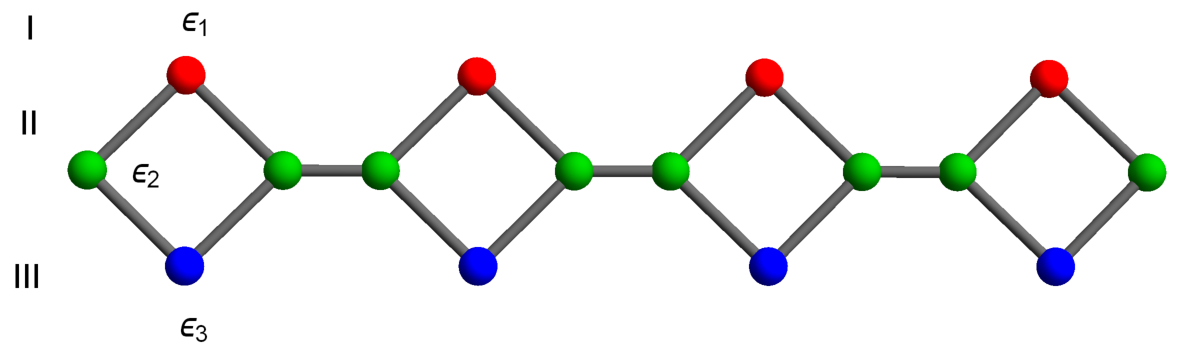}
    \caption{Schematic diagram of a diamond lattice consisting of three strands denoted by I, II, and III, subjected to three different on-site potentials $\epsilon_1$, $\epsilon_2$, and $\epsilon_3$, where $\epsilon_2$ represents the average potential of the other two components.}
    \label{fig:schematic}
\end{figure}
\subsection{Model Hamiltonian}

We consider a quasiperiodically modulated diamond lattice composed of three strands, denoted by I, II, and III, as illustrated in the schematic diagram. Each unit cell contains four lattice sites: one belonging to the upper strand (I), one to the lower strand (III), and two sites forming the middle strand (II). The tight-binding Hamiltonian describing this system can be written in the form
\begin{align}
\mathcal{H} &= 
- t \sum_{i}\sum_{m=1}^{2}
\Big(
c_{i,\mathrm{I}}^{\dagger} c_{i,\mathrm{II},m}
+ c_{i,\mathrm{III}}^{\dagger} c_{i,\mathrm{II},m}
\Big) \nonumber \\
&\quad
- t \sum_{i}\sum_{m=1}^{2}
\Big(
c_{i,\mathrm{II},m}^{\dagger} c_{i+1,\mathrm{I}}
+ c_{i,\mathrm{II},m}^{\dagger} c_{i+1,\mathrm{III}}
\Big)
+ \mathrm{H.c.} \nonumber \\
&\quad
+ \sum_{i}
\Big(
\epsilon_{1,i}\, c_{i,\mathrm{I}}^{\dagger} c_{i,\mathrm{I}}
+ \epsilon_{2,i}\sum_{m=1}^{2} 
c_{i,\mathrm{II},m}^{\dagger} c_{i,\mathrm{II},m}
+ \epsilon_{3,i}\, c_{i,\mathrm{III}}^{\dagger} c_{i,\mathrm{III}}
\Big).
\end{align}

Here $c_{i,\alpha}^{\dagger}$ ($c_{i,\alpha}$) creates (annihilates) a particle at the $i$th unit cell on strand $\alpha \in {\mathrm{I},\mathrm{II},\mathrm{III}}$, while $m=1,2$ labels the two sites on the middle strand within each diamond plaquette. The parameter $t$ denotes the nearest-neighbor inter-strand hopping amplitude, and $\epsilon_{1,i}$, $\epsilon_{2,i}$, and $\epsilon_{3,i}$ represent the site-dependent onsite potentials on strands I, II, and III, respectively. The lattice forms a diamond network in which each middle site couples to both upper and lower strands of adjacent unit cells. The upper strand is modulated as $\epsilon_{1,i}=\lambda \cos(2\pi\beta i)$, the lower strand as $\epsilon_{3,i}=\frac{\lambda}{s}\cos(2\pi\beta i)$, while the middle strand experiences their average, $\epsilon_{2,i}=(\epsilon_{1,i}+\epsilon_{3,i})/2$, where $\beta=(1+\sqrt{5})/2$.
\subsection{Localization Measures and Fractal Analysis}
To characterize the spatial nature of the eigenstates, we employ several quantitative measures that distinguish between extended, localized, and critical states. One of the most commonly used indicators is the inverse participation ratio (IPR), which evaluates the degree of spatial concentration of an eigenstate. For the $n$th eigenstate, the IPR is defined as

\begin{equation}
\mathrm{IPR}_n = \sum_{i=1}^{L} |\psi_{n,i}|^4 ,
\end{equation}

where $\psi_{n,i}$ denotes the amplitude of the eigenstate at site $i$, and $L$ represents the total number of lattice sites. In the case of completely extended states, the IPR decreases with increasing system size and scales approximately as $1/L$. In contrast, for strongly localized states the IPR remains finite even as the system size grows.

A complementary quantity is the normalized participation ratio (NPR), which provides a measure of how widely the wave function spreads over the lattice. It is defined as

\begin{equation}
\mathrm{NPR}_n = \frac{1}{L \, \mathrm{IPR}_n}.
\end{equation}

For extended states the NPR approaches values close to unity, reflecting the uniform distribution of the wave function across the system. Conversely, in the localized regime the NPR tends toward zero. Taken together, the IPR and NPR offer a consistent framework for identifying localization properties across different parameter regimes.

To obtain a global picture of the spectral behavior, these quantities are averaged over the entire set of eigenstates or within a selected energy interval near the Fermi energy. The spectral averages are defined as

\begin{equation}
\langle \mathrm{IPR} \rangle = \frac{1}{L} \sum_{n} \mathrm{IPR}_n,
\qquad
\langle \mathrm{NPR} \rangle = \frac{1}{L} \sum_{n} \mathrm{NPR}_n,
\end{equation}

where $L$ represents the number of eigenstates included in the averaging procedure. These averaged quantities provide a useful description of the overall localization tendency of the spectrum.




In addition to these measures, the multifractal nature of the eigenstates is examined through the fractal dimension. The second-order fractal dimension for the $n$th eigenstate is defined as

\begin{equation}
D_{2,n} = -\frac{\log(\mathrm{IPR}_n)}{\log L}.
\end{equation}

For extended states the fractal dimension approaches unity, whereas localized states yield values close to zero. Intermediate values indicate multifractal behavior, which is typically associated with critical states.

A spectral average of the fractal dimension is also evaluated to quantify the overall multifractal character of the system,

\begin{equation}
\langle D_2 \rangle = \frac{1}{L} \sum_{n} D_{2,n}.
\end{equation}

\section{Result and discussion}
Throughout our analysis, we consider $N=233$ plaquettes, corresponding to a total system size of $L=4N=932$ lattice sites, unless stated otherwise. The hopping amplitude is fixed at $t=1$, while the remaining parameters are specified in the respective sections where they are introduced.
\subsection{IPR and NPR resolved eigenstate as a function of disorder strength}

\begin{figure*}[t]
    \centering
   \includegraphics[width=1.0\linewidth]{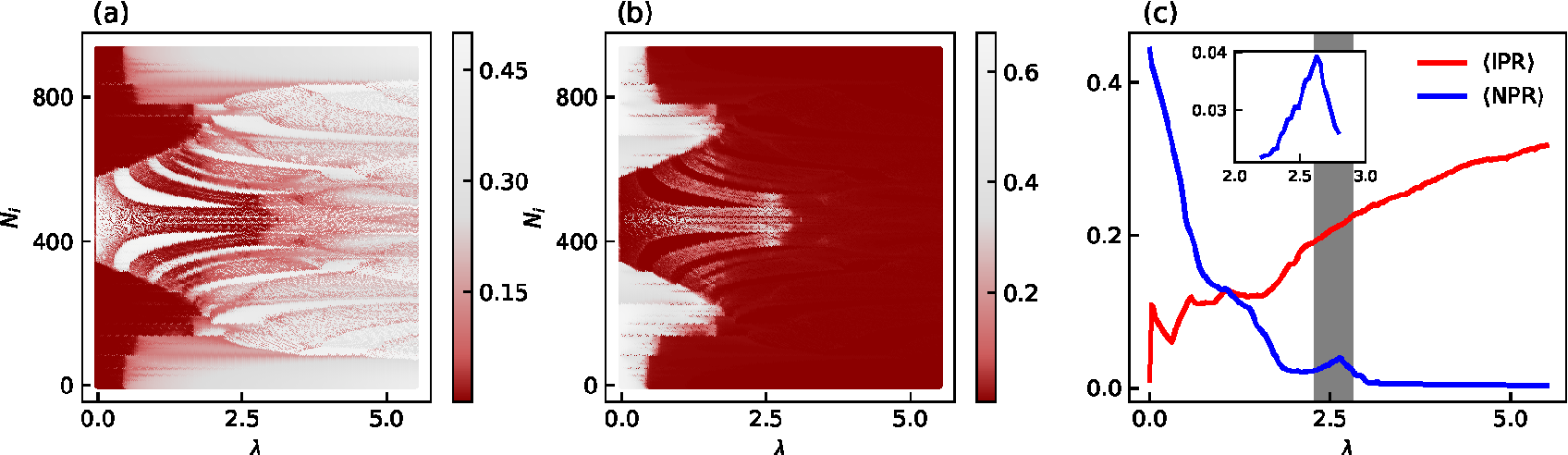}
    \caption{Eigenstates as a function of $\lambda$, colored by IPR in (a) and by NPR in (b). In (c), the averaged values $\langle IPR\rangle$ and $\langle NPR\rangle$ over the entire spectrum are plotted for $s=2$.}
    \label{fig:state1}
\end{figure*}

\begin{figure*}[t]
    \centering
   \includegraphics[width=1.0\linewidth]{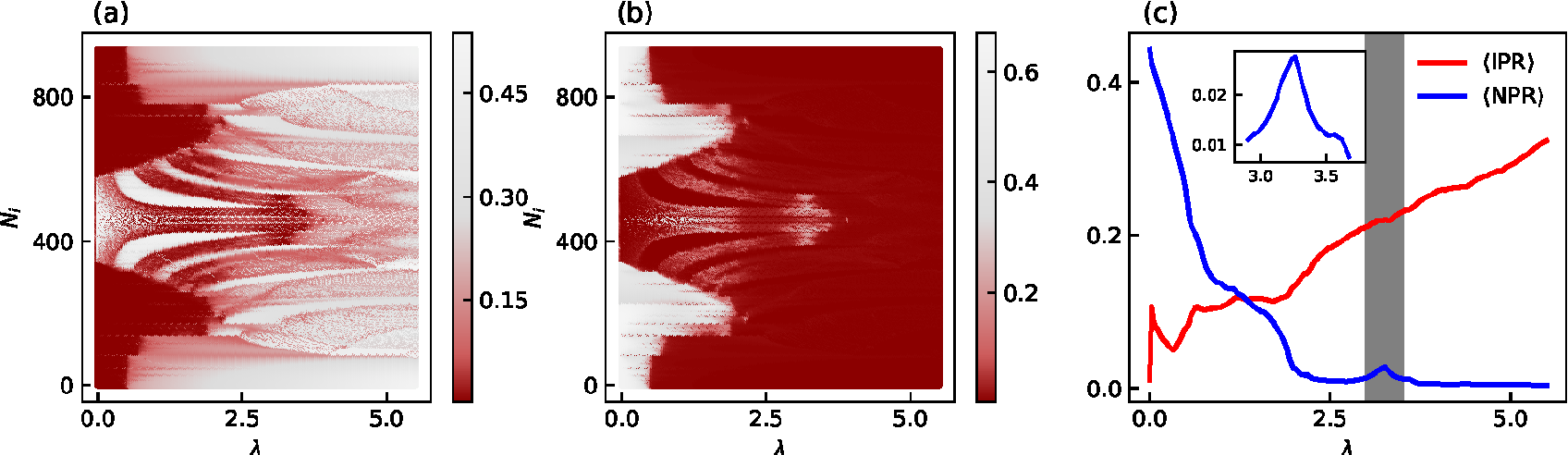}
    \caption{The same graphical layout as in Fig.~\ref{fig:state1}, but for $s=3$.}
    \label{fig:state2}
\end{figure*}

To characterize the localization properties of the one-dimensional diamond lattice with site-dependent Aubry–André–Harper (AAH) modulation, we first analyze the eigenstate-resolved IPR as a function of the modulation strength $\lambda$ for $s=2$ and $s=3$, as shown in Figs.~\ref{fig:state1}(a) and \ref{fig:state2}(a), respectively. In the color scale, low IPR values (red) correspond to extended states, while high IPR values (light color) indicate localization, with intermediate shades representing critical states. In both cases, the band-center states are already localized in the weak-disorder regime, whereas the band-edge states remain extended. As $\lambda$ increases, the band-center region exhibits alternating localization and delocalization, giving rise to a clear nonmonotonic evolution of eigenstate character. This behavior directly signals re-entrant localization–delocalization transitions. The corresponding NPR-resolved spectra [Figs.~\ref{fig:state1}(b) and \ref{fig:state2}(b)] display consistent trends, with high (low) NPR values marking extended (localized) states. Notably, the self-similar, fractal-like structure observed in both representations originates from the incommensurate nature of the AAH modulation, leading to a hierarchical spectrum with critical eigenstates. A comparison between $s=2$ and $s=3$ reveals that the re-entrant features are shifted and more pronounced in the latter, indicating that the parameter $s$ effectively controls the disorder window over which these transitions occur.

To obtain a global measure of localization, we examine the disorder dependence of the averaged quantities $\langle \mathrm{IPR} \rangle$ and $\langle \mathrm{NPR} \rangle$, shown in Figs.~\ref{fig:state1}(c) and \ref{fig:state2}(c). In the clean limit ($\lambda=0$), finite values of $\langle \mathrm{IPR} \rangle \sim 0.1$ and $\langle \mathrm{NPR} \rangle \sim 0.4$ indicate the coexistence of localized and extended states, reflecting the intrinsic quasiperiodic and geometrical effects of the system. With increasing $\lambda$, both quantities exhibit pronounced oscillatory behavior characterized by multiple peaks and dips, corresponding to successive localization–delocalization transitions across the spectrum. In particular, the enhancement of $\langle \mathrm{NPR} \rangle$ following its prior suppression (around $\lambda \approx 2.5$ for $s=2$ and $\lambda \approx 3$ for $s=3$), as highlighted by the shaded regions and insets, provides clear evidence of re-entrant delocalization. The stronger and sharper oscillations observed for $s=3$ further demonstrate that increasing $s$ enhances the re-entrant response, establishing it as an effective control parameter for tuning the interplay between quasiperiodicity and disorder.

\begin{figure*}[t]
    \centering
   \includegraphics[width=1.0\linewidth]{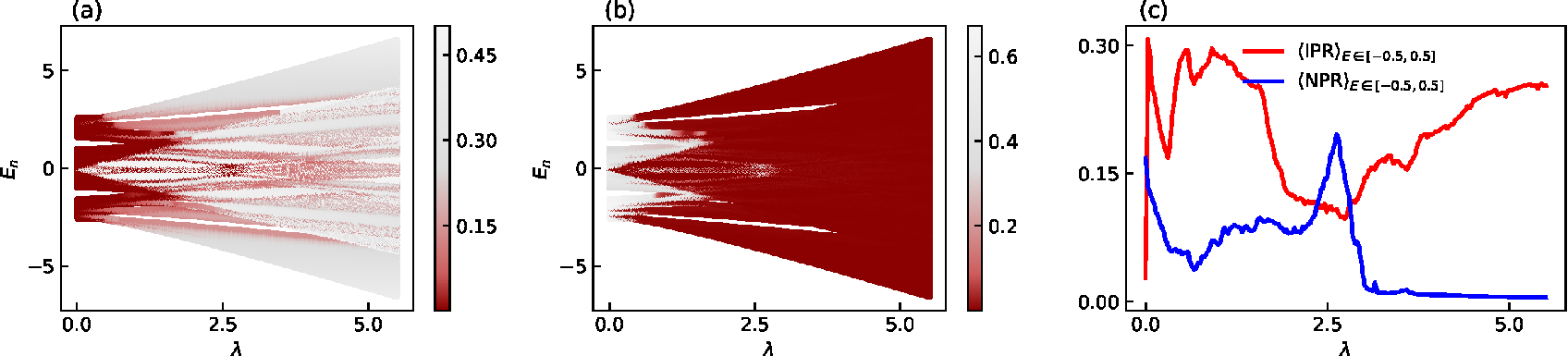}
    \caption{Energy eigenvalues as a function of $\lambda$, colored by IPR in (a) and NPR in (b), respectively. The averaged values $\langle IPR\rangle$ and $\langle NPR\rangle$ for eigenvalues within the range $-0.5$ to $0.5$ are presented in (c). All calculations are performed for $s=2$.}
    \label{fig:band1}
\end{figure*}

\begin{figure*}[t]
    \centering
   \includegraphics[width=1.0\linewidth]{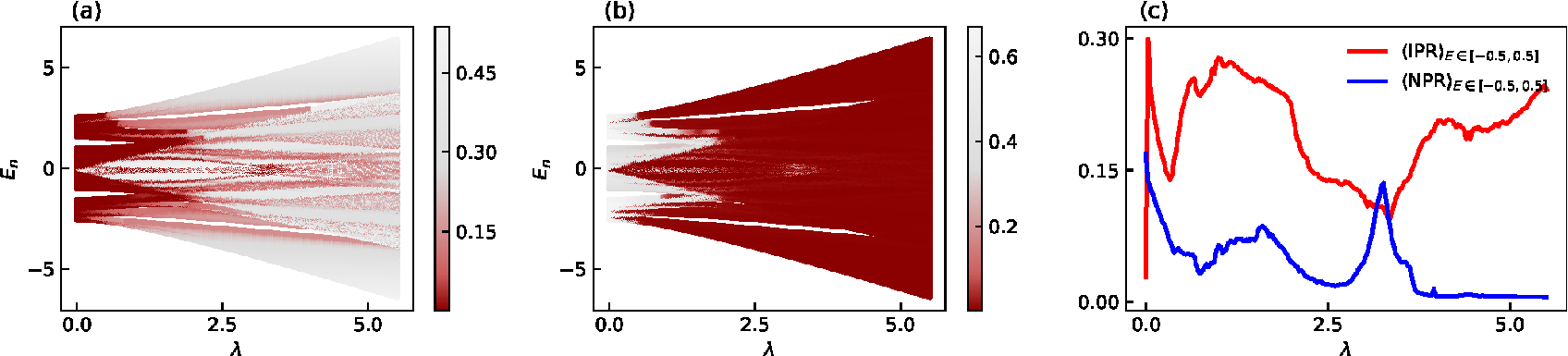}
    \caption{Similar characteristics as in Fig.~\ref{fig:band1}, but for $s=3$.}
    \label{fig:band2}
\end{figure*}

\subsection{Energy spectrum characterized by IPR and NPR with varying $\lambda$}
\begin{figure*}[t]
    \centering
   \includegraphics[width=1.0\linewidth]{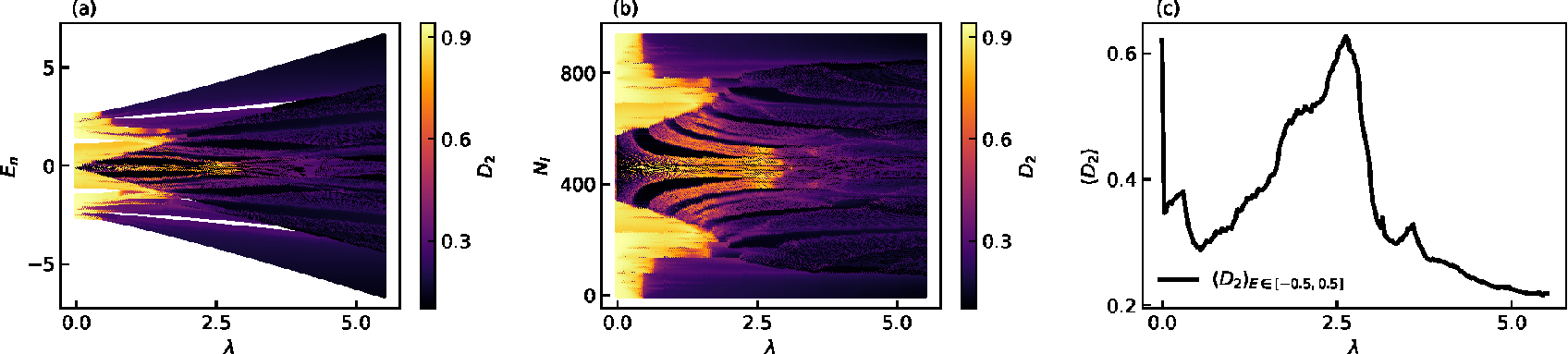}
    \caption{Fractal dimension $D_2$ is represented in the eigenvalue spectrum in (a) and in the eigenstate spectrum in (b). In (c), the averaged $D_2$ is plotted as a function of disorder strength for the energy window $-0.5$ to $0.5$. All calculations are performed for $s=2$.}
    \label{fig:d211}
\end{figure*}

\begin{figure*}[t]
    \centering
   \includegraphics[width=1.0\linewidth]{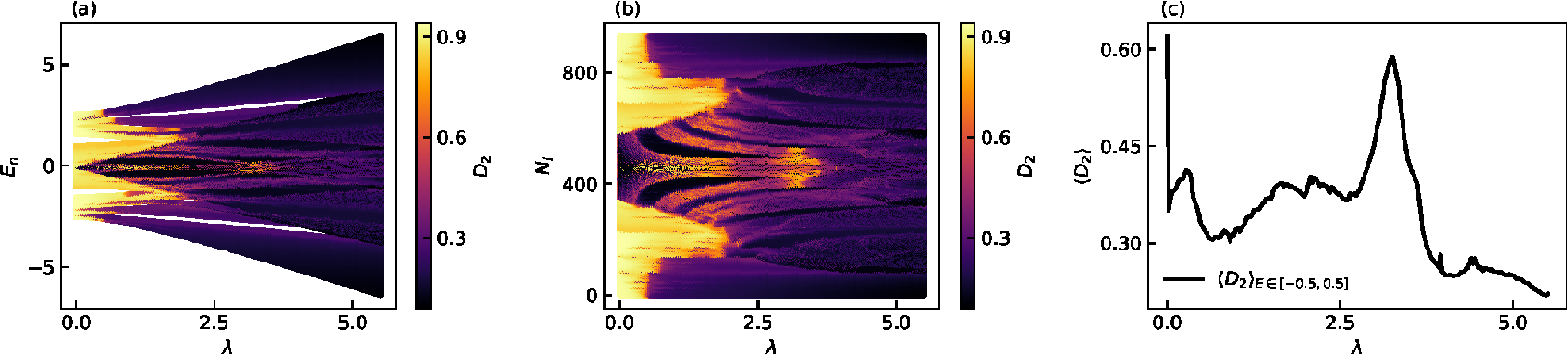}
    \caption{Similar layout like Fig.~\ref{fig:d211} but for $s=3$.}
    \label{fig:d212}
\end{figure*}

We next analyze the eigenvalue spectrum resolved by IPR and NPR as a function of disorder strength $\lambda$ for both $s=2$ and $s=3$, as shown in Figs.~\ref{fig:band1} and \ref{fig:band2}. In the nearly clean limit ($\lambda \approx 0$), the majority of eigenstates exhibit vanishingly small IPR values, indicating extended (conducting) behavior, except for the states located at the band center, which remain localized due to the intrinsic geometry of the diamond lattice. As $\lambda$ increases, these band-center states display pronounced sensitivity and undergo multiple transitions between extended and localized regimes, even for small variations in $\lambda$. This clearly establishes the modulation strength as an effective control parameter for tuning repeated localization–delocalization transitions. The presence of intermediate color regions further indicates partially localized (critical) states that persist over a broad range of $\lambda$, demonstrating that the system supports a robust mixed phase rather than a sharp separation between conducting and insulating regimes. The corresponding NPR-resolved spectra [Figs.~\ref{fig:band1}(b) and \ref{fig:band2}(b)] exhibit consistent behavior, where alternating regions of high and low NPR values confirm multiple re-entrant localization–delocalization transitions. A comparison between $s=2$ and $s=3$ reveals that the characteristic $\lambda$ windows associated with these transitions are shifted and broadened for $s=3$, highlighting the role of $s$ in tuning the disorder range over which re-entrant behavior occurs.

To further quantify these effects, we examine the variation of the averaged quantities $\langle \mathrm{IPR} \rangle$ and $\langle \mathrm{NPR} \rangle$ at the band center as a function of $\lambda$, as shown in Figs.~\ref{fig:band1}(c) and \ref{fig:band2}(c). In the weak-disorder regime, $\langle \mathrm{IPR} \rangle$ remains close to zero, indicating extended behavior; however, even a small increase in $\lambda$ leads to a rapid rise (up to $\sim 0.3$), signaling a sharp onset of localization. With further increase in $\lambda$, both $\langle \mathrm{IPR} \rangle$ and $\langle \mathrm{NPR} \rangle$ exhibit pronounced nonmonotonic behavior characterized by multiple peaks and dips, reflecting successive localization–delocalization transitions and the emergence of a critical regime. Notably, around $\lambda \approx 2.8$ for $s=2$ and $\lambda \approx 3$ for $s=3$, $\langle \mathrm{NPR} \rangle$ shows a marked enhancement accompanied by a dip in $\langle \mathrm{IPR} \rangle$, indicating a re-entrant transition to an extended phase. At larger $\lambda$, the eigenstates relocalize, completing the sequence of re-entrant transitions. The stronger and broader oscillatory features observed for $s=3$ further demonstrate that increasing $s$ enhances and shifts the re-entrant response, confirming its role as an effective tuning parameter for controlling localization properties in the system.

\begin{figure}[t]
    \centering
   \includegraphics[width=1.0\linewidth]{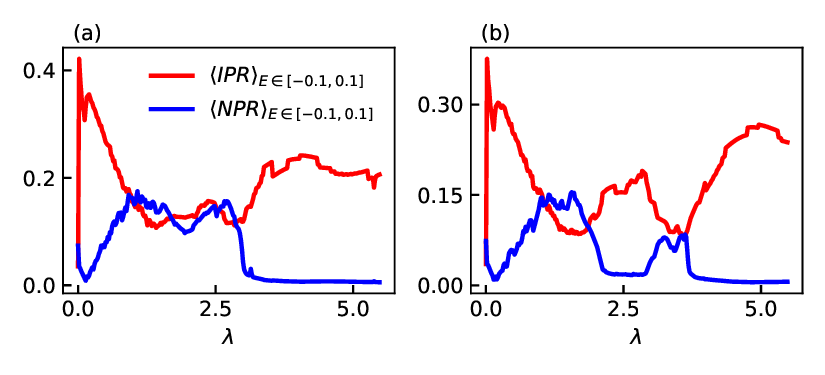}
    \caption{The averaged IPR and NPR for eigenenergies within the range $-0.1$ to $0.1$, for $s=2$ in (a) and $s=3$ in (b), respectively.}
    \label{fig:d22}
\end{figure}
\subsection{Eigenstate Characterization via Fractal Dimension and Re-entrant Transitions}
To further characterize the nature of the eigenstates, we analyze the density plots of the fractal dimension $D_n$, where $D_n \rightarrow 1$ ($0$) corresponds to fully extended (localized) states, while intermediate values indicate critical multifractal behavior. Figures~\ref{fig:d211}(a) and \ref{fig:d212}(a) show the eigenenergies as a function of disorder strength $\lambda$, color-coded by $D_n$ for $s=2$ and $s=3$, respectively. In both cases, the band-center states exhibit low $D_n$ values in the weak-disorder regime, indicating localization, while increasing $\lambda$ drives them toward higher $D_n$, signaling enhanced delocalization. The presence of intermediate color regions confirms the existence of critical states over a broad range of $\lambda$. The alternating evolution between low and high $D_n$ values with increasing disorder strength provides clear evidence of re-entrant localization–delocalization transitions. This behavior is further illustrated in Figs.~\ref{fig:d211}(b) and \ref{fig:d212}(b), which display $D_n$ as a function of eigenstate index and $\lambda$. In both cases, the band-center states undergo repeated transitions between localized, extended, and critical regimes, while the self-similar, fractal-like patterns reflect the incommensurate AAH modulation that generates hierarchical spectral structures and multifractal eigenstates. Notably, compared to $s=2$, the re-entrant features for $s=3$ are shifted and more pronounced, particularly in the range $\lambda \sim 3$–$3.5$, indicating an enhanced sensitivity to quasiperiodic modulation.

We further examine the disorder dependence of the averaged fractal dimension $D_n$, as shown in Figs.~\ref{fig:d211}(c) and \ref{fig:d212}(c) for $s=2$ and $s=3$, respectively. In the clean limit, $D_n \approx 0.6$, indicating predominantly extended character. With increasing $\lambda$, $D_n$ initially decreases, marking the onset of localization, followed by a sequence of nonmonotonic oscillations characterized by peaks and dips that reflect repeated transitions between extended and localized regimes. For $s=2$, these oscillations appear over a relatively broad range of $\lambda$, while for $s=3$ a more pronounced enhancement is observed, including a distinct peak around $\lambda \sim 3.5$ that signals re-entrant delocalization. The persistence of intermediate $D_n$ values further confirms the presence of a critical regime. Overall, the strong nonmonotonic variation of $D_n$ with $\lambda$ highlights the interplay between quasiperiodicity and disorder, and establishes the parameter $s$ as an effective control knob for tuning the extent and location of re-entrant localization–delocalization transitions.

\begin{figure}[t]
    \centering
   \includegraphics[width=1.0\linewidth]{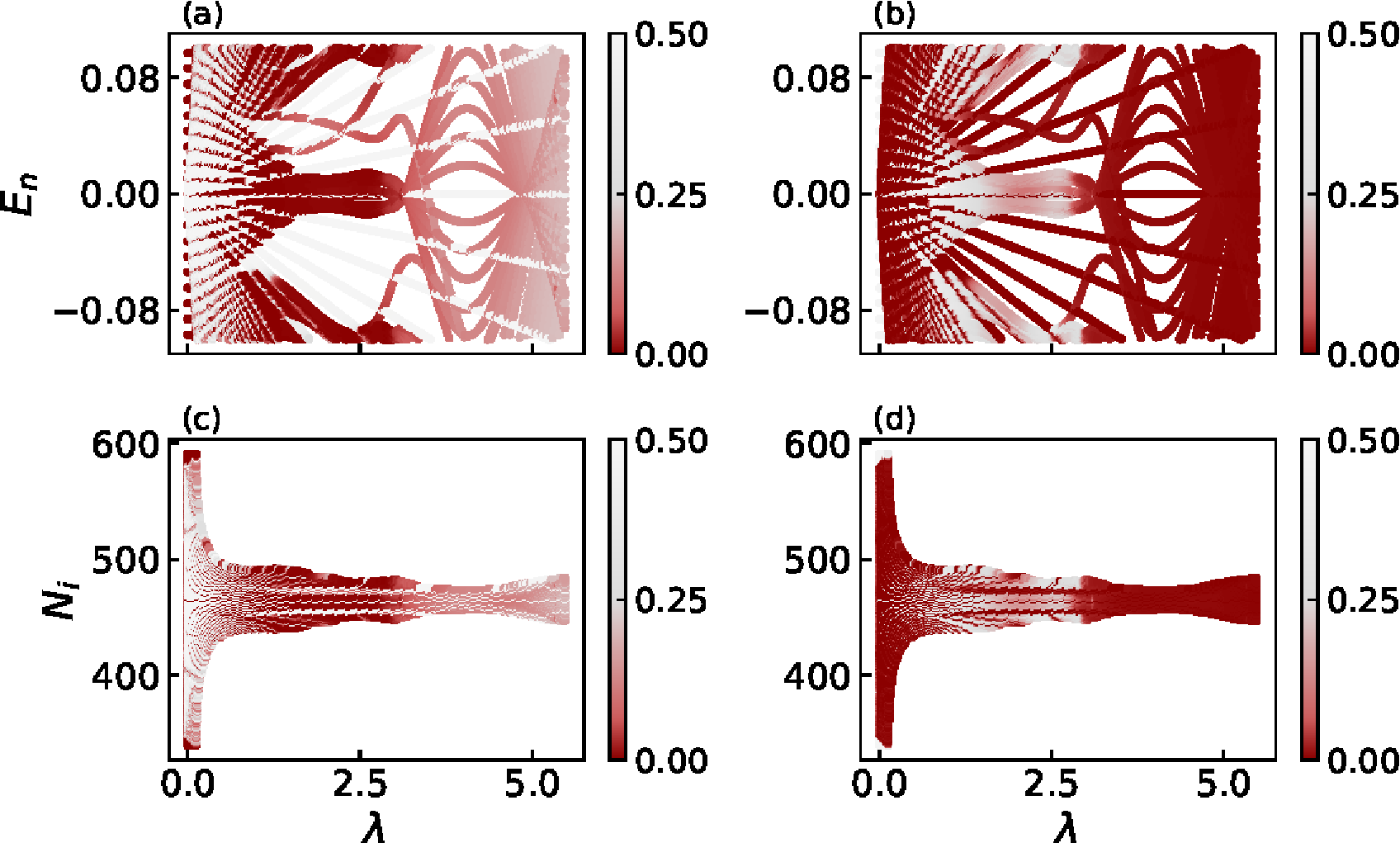}
    \caption{Energy eigenvalues in (a) and (b), and the corresponding eigenstates in (c) and (d), plotted as a function of $\lambda$. The first column is colored by IPR and the second column by NPR. The results are shown for the energy window $-0.1$ to $0.1$ and the corresponding states for $s=2$.}
    \label{fig:d23}
\end{figure}

\begin{figure}[t]
    \centering
   \includegraphics[width=1.0\linewidth]{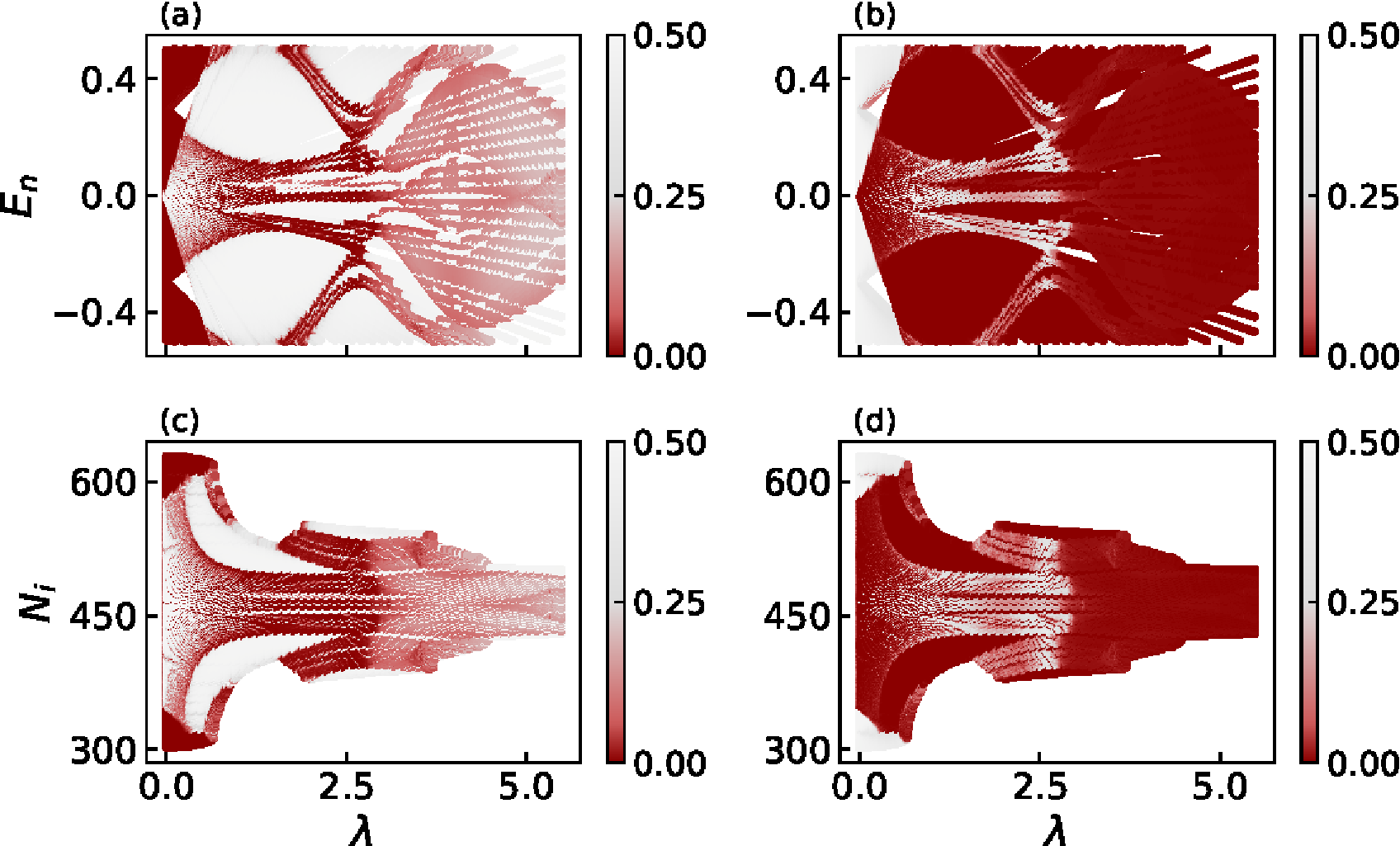}
    \caption{Same layout as Fig.~\ref{fig:d23}, but for the energy window $-0.5$ to $0.5$.}
    \label{fig:d24}
\end{figure}

\begin{figure}[t]
    \centering
   \includegraphics[width=1.0\linewidth]{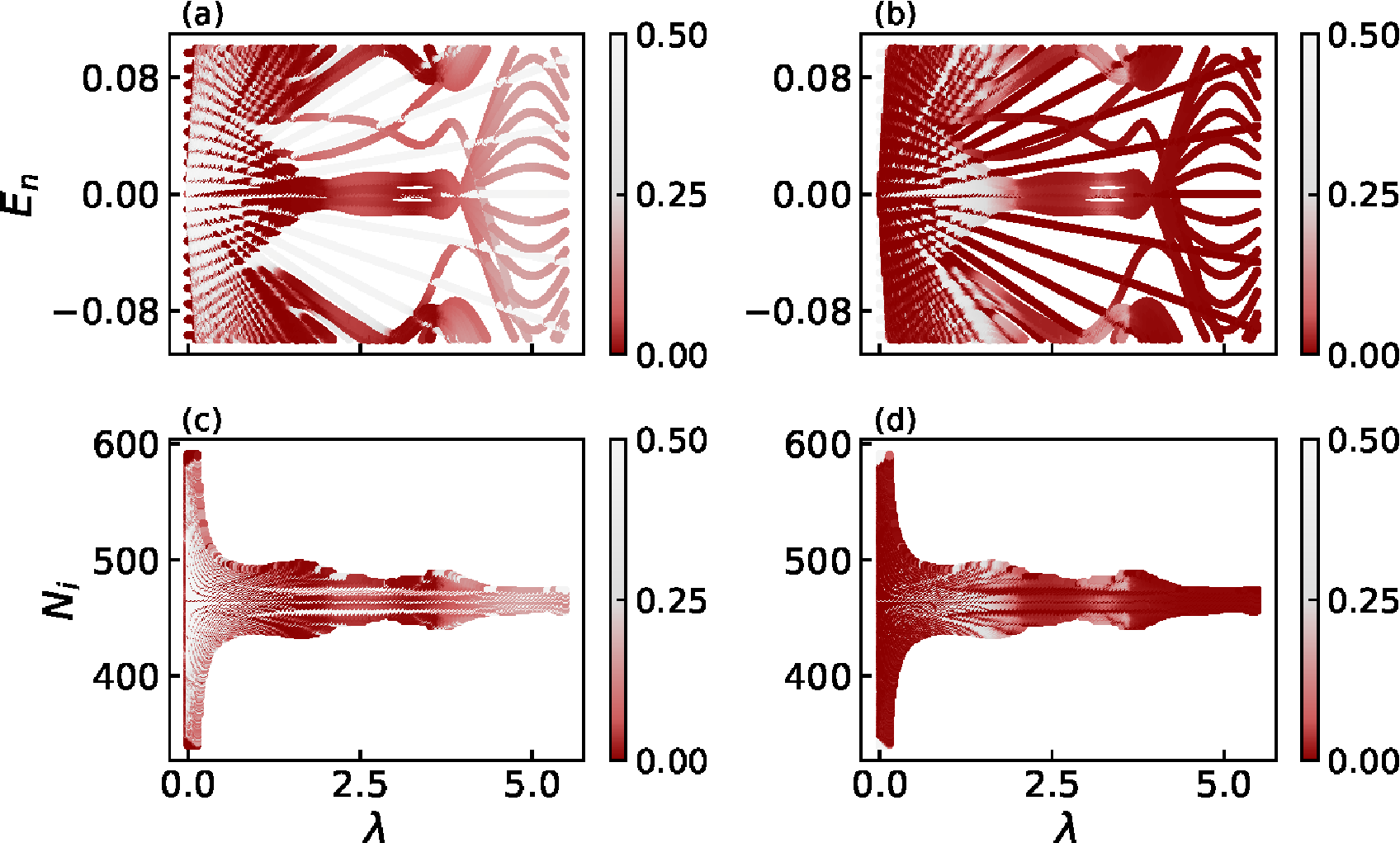}
    \caption{Same layout as Fig.~\ref{fig:d23}, but for $s=3$.}
    \label{fig:d25}
\end{figure}
\subsection{Detailed analysis of the band-center region}

To obtain a sharper picture of the re-entrant localization behavior, we focus on the band-center region and analyze the magnified energy window $-0.1~\mathrm{eV} \le E \le 0.1~\mathrm{eV}$. Figures~\ref{fig:d22}(a) and \ref{fig:d22}(b) show the variation of $\langle IPR \rangle$ and $\langle NPR \rangle$ for $s=2$ and $s=3$, respectively. In both cases, the averaged quantities exhibit pronounced nonmonotonic dependence on $\lambda$, marked by alternating peaks and dips that signal repeated localization–delocalization transitions. Notably, for $s=3$ the oscillations become significantly sharper, indicating a stronger and more well-defined re-entrant response. 
\begin{figure}[t]
    \centering
   \includegraphics[width=1.0\linewidth]{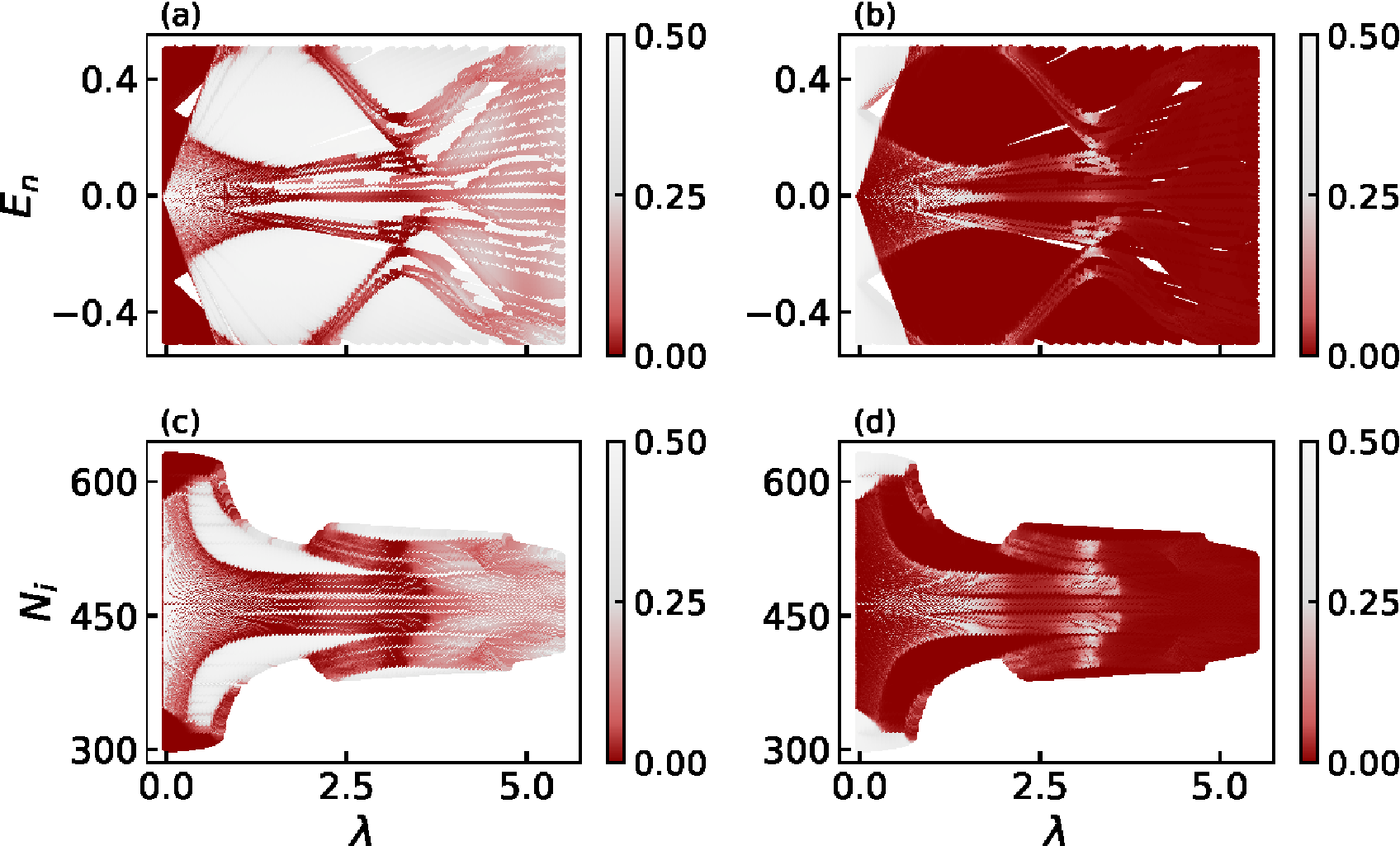}
    \caption{Same layout as Fig.~\ref{fig:d24}, but for $s=3$.}
    \label{fig:d26}
\end{figure}
To resolve the microscopic origin of this behavior, we examine the spectral evolution color coded by IPR and NPR (Figs.~\ref{fig:d23}–\ref{fig:d26}). In the clean limit (Figs.~\ref{fig:d23}–\ref{fig:d24}), the states are predominantly extended, while increasing $\lambda$ drives them toward localization. However, within an intermediate range (around $\lambda \sim 0.3$), several states regain extended character before localizing again at higher disorder strengths, producing a sequence of alternating conducting and insulating regions. This structure is further corroborated by tracking band-center states (indices $\sim 300$–$600$), which clearly evolve from localized to extended and back to localized regimes with increasing $\lambda$. Importantly, this behavior persists over both narrow and wide energy windows, demonstrating that the re-entrant transitions are not confined to a specific spectral region but are a global feature of the system.
\begin{figure}[t]
    \centering
   \includegraphics[width=0.8\linewidth]{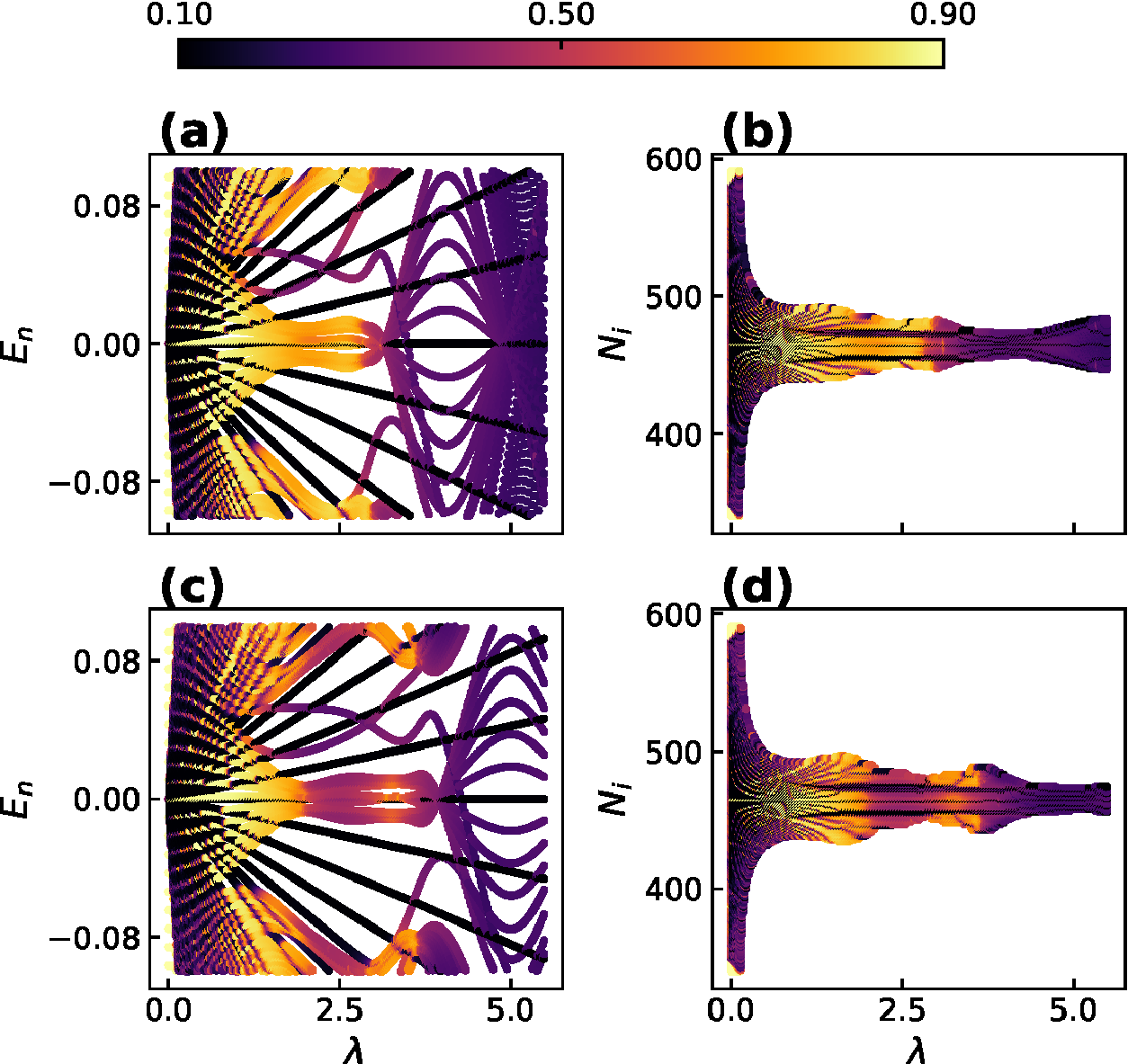}
    \caption{Energy eigenvalues and the corresponding eigenstates colored by $D_2$, with the top row for $s=2$ and the bottom row for $s=3$, respectively, for the energy range $-0.1$ to $0.1$.}
    \label{fig:d27}
\end{figure}
A similar analysis for $s=3$ (Figs.~\ref{fig:d25} and \ref{fig:d26}) reveals an even more pronounced re-entrant pattern, with an extended region spanning a broader range of $\lambda$ (approximately $\lambda \sim 0.3$–$4.0$). This clearly indicates that the parameter $s$ effectively controls the width and location of the conducting and insulating phases, acting as a tuning knob for the re-entrant behavior.

\begin{figure}[t]
    \centering
   \includegraphics[width=0.8\linewidth]{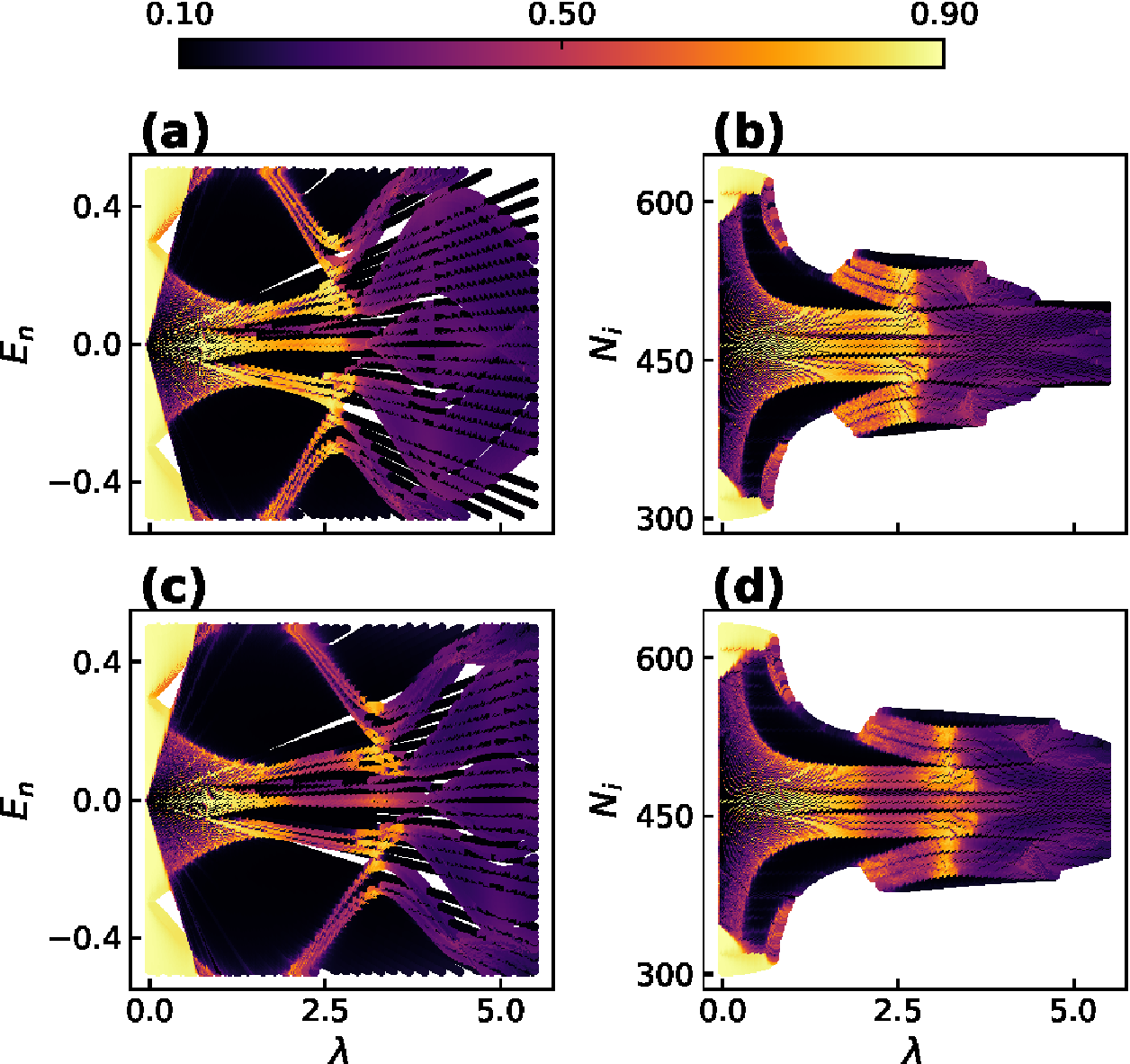}
    \caption{Same as Fig.~\ref{fig:d27}, but for the energy range $-0.5$ to $0.5$.}
    \label{fig:d28}
\end{figure}

To further characterize the nature of the eigenstates, we analyze the fractal dimension $D_2$, as shown in Figs.~\ref{fig:d27} and \ref{fig:d28}. The color-coded spectra reveal clear alternation between localized ($D_2 \rightarrow 0$), extended ($D_2 \rightarrow 1$), and intermediate critical regimes, providing direct evidence of multifractality in the system. In particular, for $s=3$ a broad mixed-phase region emerges around $\lambda \approx 2.5$, which is significantly narrower for $s=2$, highlighting the enhanced role of quasiperiodic modulation. The corresponding eigenstate-resolved plots near the band center further confirm repeated transitions between different regimes as $\lambda$ increases.

\begin{figure}[t]
    \centering
   \includegraphics[width=1.0\linewidth]{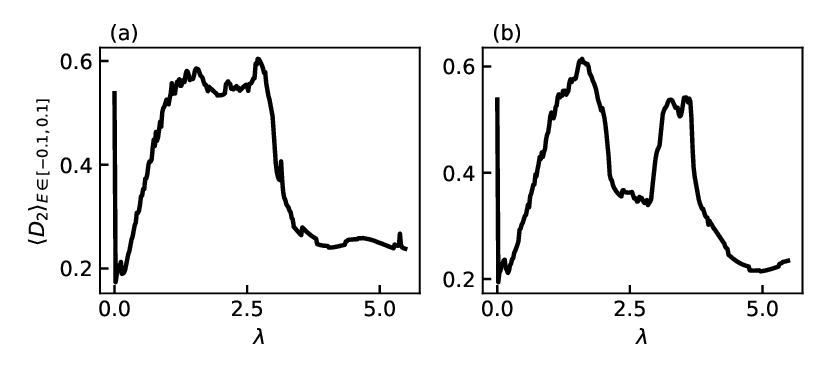}
    \caption{Average $D_2$ as a function of $\lambda$, calculated from the states within the energy range $-0.1$ to $0.1$, for $s=2$ and $s=3$ in (a) and (b), respectively.}
    \label{fig:d29}
\end{figure}

Finally, the averaged fractal dimension $D_2$ (Fig.~\ref{fig:d29}) exhibits a sequence of oscillations with $\lambda$, closely mirroring the behavior of $\langle IPR \rangle$ and $\langle NPR \rangle$. The enhanced amplitude of these oscillations for $s=3$ reinforces the conclusion that increasing $s$ amplifies the re-entrant localization–delocalization behavior. Taken together, these results establish that the observed re-entrant transitions are robust, multifractal in nature, and tunable via the system parameters, providing a unified understanding of the interplay between quasiperiodicity and disorder.

\begin{figure}[t]
    \centering
   \includegraphics[width=1.0\linewidth]{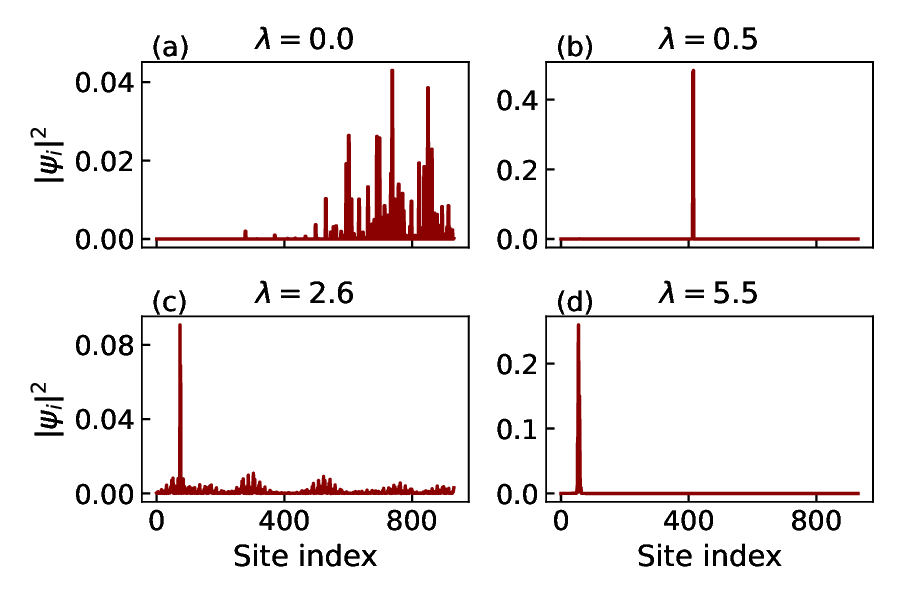}
    \caption{Probability density as a function of site index for the $451^{\mathrm{th}}$ eigenstate at $s=2$.}
    \label{fig:d30}
\end{figure}
\subsection{Probability density with site index for different values of $\lambda$}


\begin{figure}[t]
    \centering
   \includegraphics[width=1.0\linewidth]{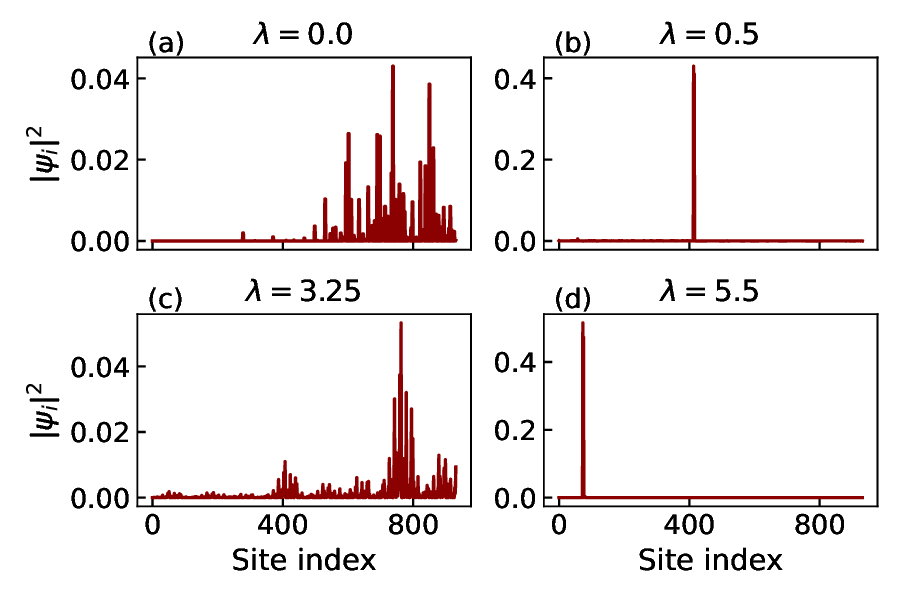}
    \caption{Same layout as Fig.~\ref{fig:d30}, but for $s=3$. }
    \label{fig:d31}
\end{figure}
To gain direct real-space insight into the re-entrant localization–delocalization behavior, we track the evolution of a representative eigenstate ($n=451$) with increasing disorder strength $\lambda$ for both $s=2$ and $s=3$, as shown in Figs.~\ref{fig:d30} and \ref{fig:d31}, respectively. In the clean limit ($\lambda=0$), the wave function is distributed across the lattice, indicating an extended state. With increasing $\lambda$ (e.g., $\lambda=0.5$), the state becomes spatially confined, signaling localization. Remarkably, upon further increasing $\lambda$ (around $\lambda \approx 2.6$ for $s=2$ and $\lambda \approx 3.25$ for $s=3$), the eigenstate regains its extended character, before eventually localizing again at larger disorder strengths ($\lambda=5.5$). This sequence of alternating spatial profiles, extended to localized, followed by re-emergent delocalization and subsequent relocalization, provides clear real-space evidence of re-entrant behavior. Moreover, the shift in the re-entrant window between $s=2$ and $s=3$ highlights the role of $s$ as an effective tuning parameter controlling the onset and extent of the conducting regime. These observations are in full agreement with the spectral signatures discussed earlier, thereby establishing a consistent and unified picture of disorder-driven re-entrant transitions in the system.
\subsection{IPR and NPR with eigenstate index for fixed $\lambda$}
\begin{figure}[t]
    \centering
   \includegraphics[width=1.0\linewidth]{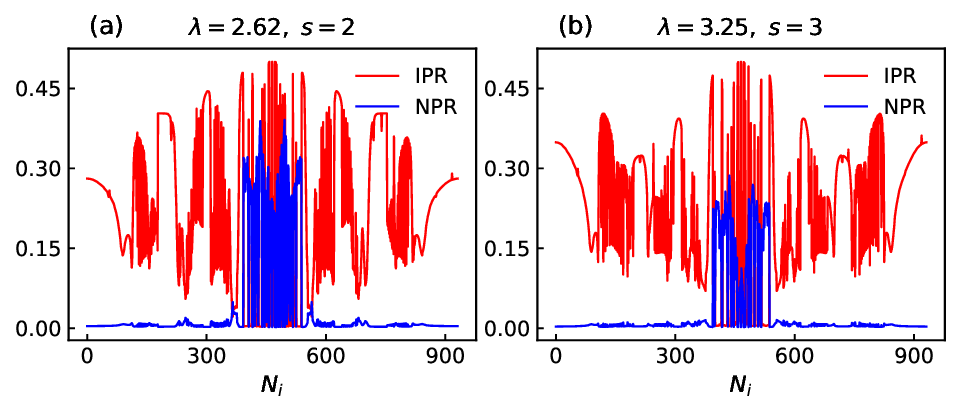}
    \caption{ IPR and NPR with eigenstates}
    \label{fig:butterfly}
\end{figure}

To further examine the reentrant regime, we analyze the variation of the IPR and NPR of the eigenstates for two representative values of the quasiperiodic strength $\lambda$, where the reentrant behavior is most prominent (see Fig.~\ref{fig:state2}). In Fig.~\ref{fig:butterfly}(a), we fix $\lambda = 2.62$ for $s=3$, while in Fig.~\ref{fig:butterfly}(b) we consider $\lambda = 3.25$ for the same value of $s$. 
Within this reentrant region, the IPR exhibits an oscillatory behavior across the spectrum, indicating strong variations in the localization characteristics of the eigenstates. In contrast, the NPR remains generally small and shows only weak peaks over most parts of the spectrum. However, in the central energy region the NPR displays comparatively more pronounced peaks, suggesting that the corresponding eigenstates undergo a noticeable transition between extended and localized character at these particular disorder strengths. This behavior further supports the presence of reentrant localization in the system.

\subsection{Counting of extended states}

\begin{figure*}[t]
    \centering
   \includegraphics[width=0.7\linewidth]{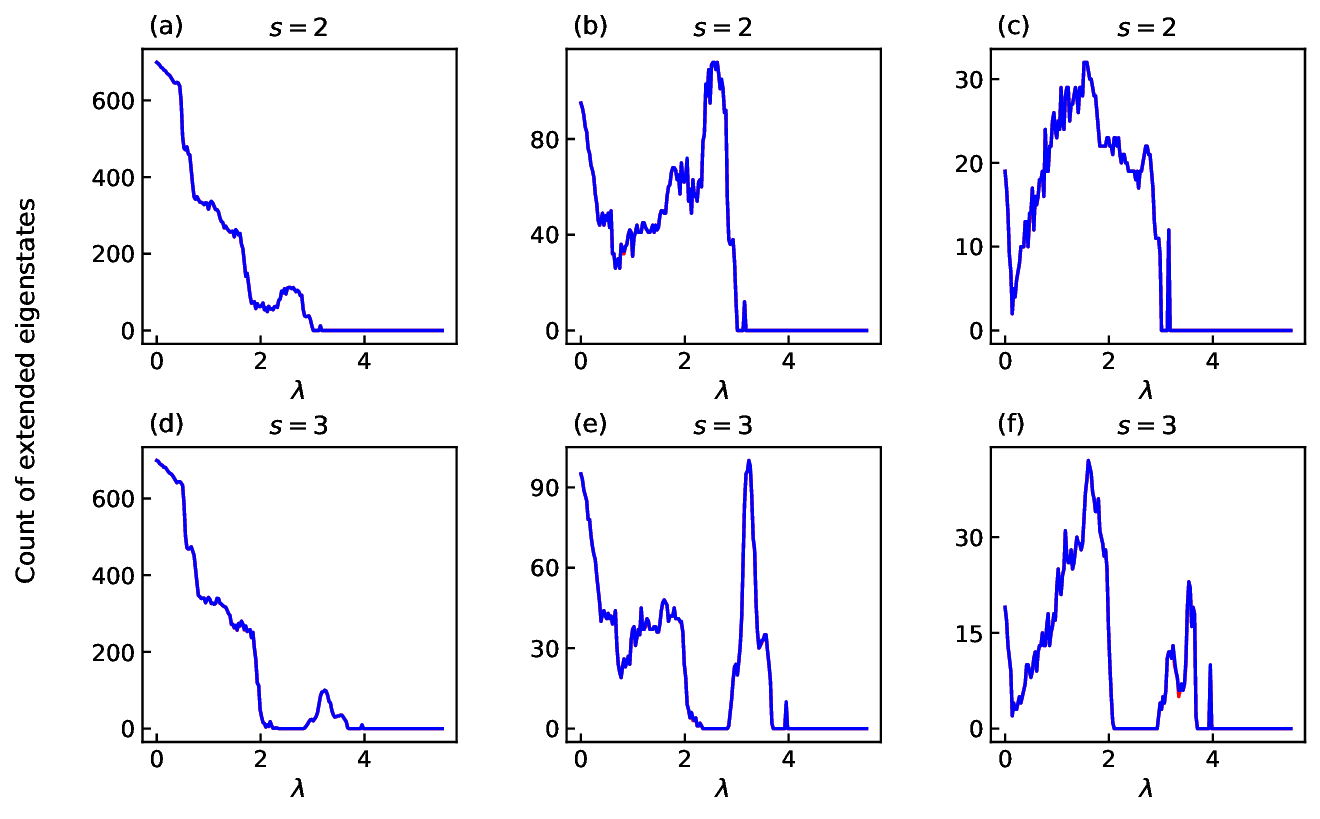}
    \caption{Count of extended eigenstates with $IPR \leq 0.01$ as a function of $\lambda$. The first column corresponds to the full energy spectrum, while the second and third columns represent eigenstates associated with energy ranges $[-0.5,0.5]$ and $[-0.1,0.1]$, respectively.}
    \label{fig:count1}
\end{figure*}

Finally, we analyze how the number of extended eigenstates varies with the quasiperiodic strength $\lambda$, as shown in Fig.~\ref{fig:count1}. Fig.~\ref{fig:count1}(a) presents the variation of the total number of extended eigenstates with $\lambda$ for $s=2$. For small $\lambda$, the system hosts a relatively large number of extended states, which gradually decreases as $\lambda$ increases from $0$ to $2$. Around $\lambda \approx 2$, the number of extended states reaches a minimum. Interestingly, upon further increasing $\lambda$, the number of extended states increases again before eventually diminishing for $\lambda \gtrsim 3$. This nonmonotonic behavior indicates a reentrant delocalization, where extended states reappear after an initially localized regime. A similar trend is observed for $s=3$, as shown in Fig.~\ref{fig:count1}(d), where the feature becomes even more pronounced. In this case, the number of extended states nearly vanishes before rising again to form a bell-shaped structure in the range $\lambda \approx 2.8$ to $4$, providing a clear signature of reentrant behavior.
To gain further insight, we next focus on the central part of the spectrum, where the reentrant phenomenon was previously found to be most prominent. In Fig.~\ref{fig:count1}(b), we plot the number of extended eigenstates as a function of $\lambda$ for $s=2$ within the energy window $-0.5~\mathrm{eV} \leq E \leq 0.5~\mathrm{eV}$. At $\lambda = 0$, the number of extended states is approximately $90$, which gradually decreases with increasing $\lambda$, accompanied by several small peaks and dips. Beyond $\lambda \approx 1$, the curve shows a gradual increase and exhibits a small peak near $\lambda \approx 2$. Notably, around $\lambda \approx 2.8$, the number of extended states rises sharply to a higher peak of approximately $110$, exceeding the value at $\lambda=0$. This behavior provides strong evidence for reentrant delocalization. A similar feature is observed for $s=3$ in Fig.~\ref{fig:count1}(e), although the characteristic range of $\lambda$ shifts slightly: the number of extended states drops sharply near $\lambda \approx 2$ and then increases again around $\lambda \approx 3.2$, consistent with the reentrant delocalization scenario.

Finally, we examine an even narrower energy window around the band center, $-0.1~\mathrm{eV} \leq E \leq 0.1~\mathrm{eV}$. As shown in Fig.~\ref{fig:count1}(c) for $s=2$, the number of extended states initially increases with $\lambda$, reaches a peak near $\lambda \approx 2$, and subsequently decreases with multiple smaller peaks and dips. For $s=3$, shown in Fig.~\ref{fig:count1}(f), a similar trend is observed: the number of extended states grows with increasing $\lambda$, peaks around $\lambda \approx 2$, and then decreases sharply before exhibiting additional peaks in the range $\lambda \approx 3$ to $4$. These results demonstrate that the reentrant delocalization behavior persists even within a narrow central energy window, further confirming the robustness of the phenomenon in the present system.

\begin{figure}[t]
    \centering
   \includegraphics[width=1.0\linewidth]{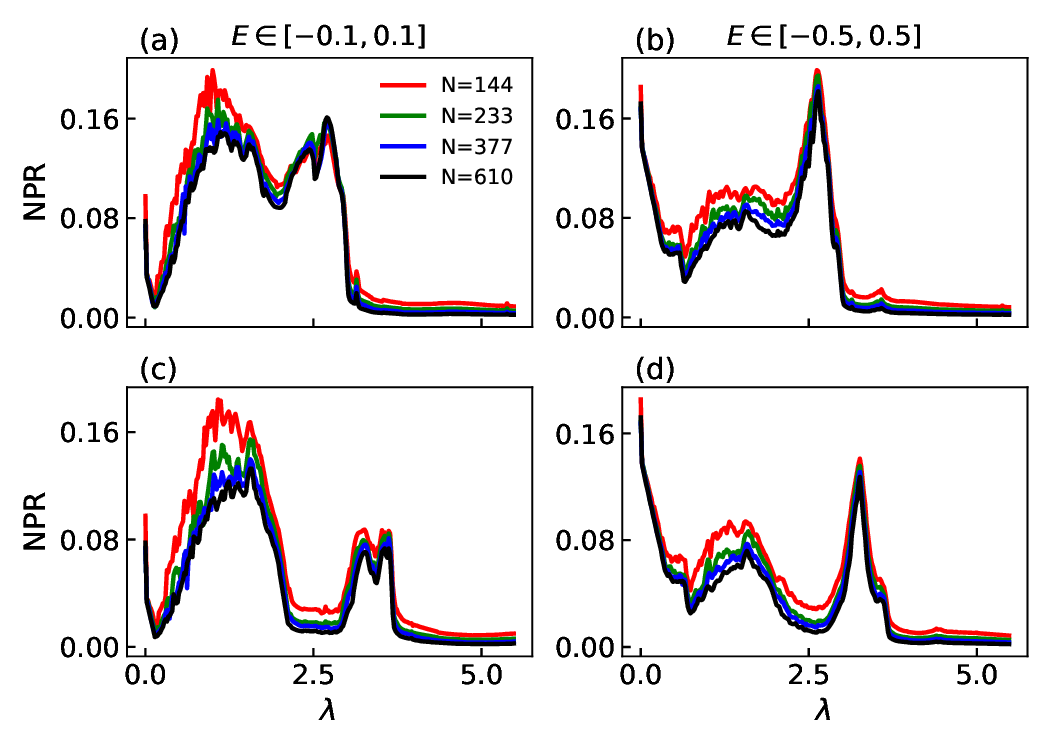}
    \caption{Average NPR with $\lambda$ for the energy ranges $-0.1$ to $0.1$ (first column) and $-0.5$ to $0.5$ (second column) for four different block sizes, as shown in the figure. The top row corresponds to $s=2$, while the bottom row corresponds to $s=3$, respectively.}
    \label{fig:d32}
\end{figure}
\begin{figure}[t]
    \centering
   \includegraphics[width=1.0\linewidth]{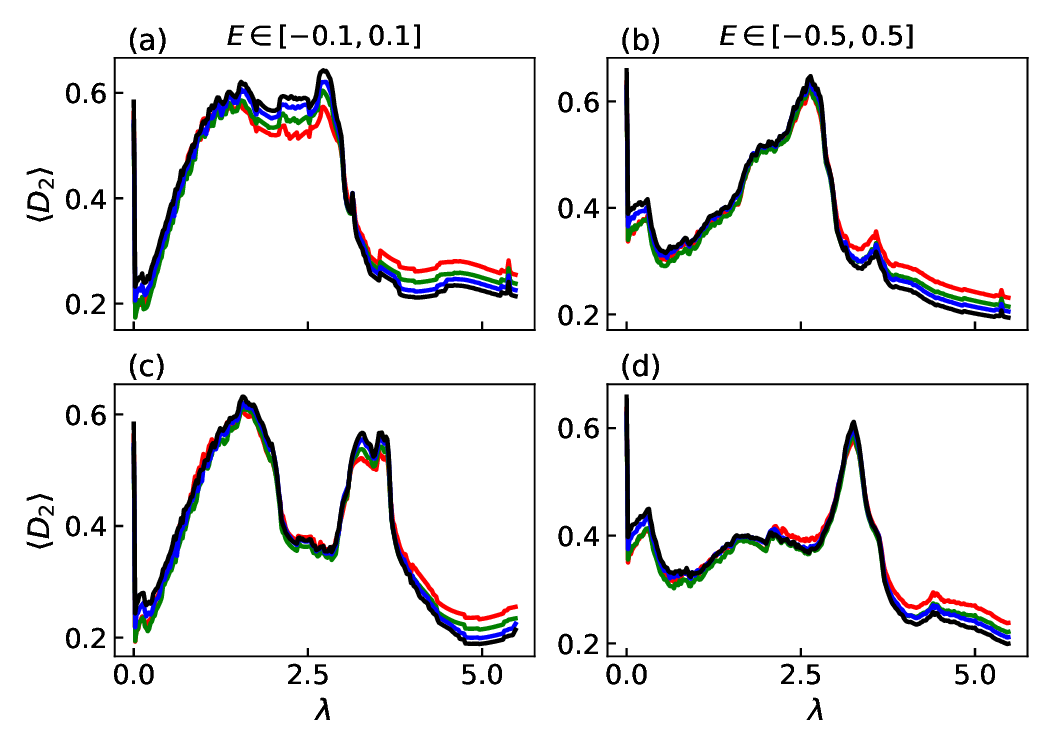}
    \caption{Same layout as Fig.~\ref{fig:d32}, but with the average $D_2$ plotted instead of the average NPR.}
    \label{fig:d33}
\end{figure}
\subsection{System size dependence of $\langle NPR\rangle$ and $\langle D_2 \rangle$ with $\lambda$}
We first examine the system-size dependence of the re-entrant behavior through the averaged normalized participation ratio, $\langle \mathrm{NPR} \rangle$, for plaquette sizes $N=144$, $233$, $377$, and $610$. Figure~\ref{fig:d32}(a) shows $\langle \mathrm{NPR} \rangle$ as a function of disorder strength $\lambda$ for $s=2$ within the energy window $-0.1~\mathrm{eV} \le E \le 0.1~\mathrm{eV}$. As $\lambda$ increases, $\langle \mathrm{NPR} \rangle$ exhibits a sequence of alternating maxima and minima, indicating repeated transitions between extended and localized regimes. This behavior persists over the wider energy range $-0.5~\mathrm{eV} \le E \le 0.5~\mathrm{eV}$, as shown in Fig.~\ref{fig:d32}(b), demonstrating that the observed features are not restricted to a narrow spectral window. The corresponding results for $s=3$ are presented in Figs.~\ref{fig:d32}(c) and \ref{fig:d32}(d), for the narrow and wide energy windows, respectively. In both cases, the oscillatory variation of $\langle \mathrm{NPR} \rangle$ with $\lambda$ remains clearly visible. Importantly, the persistence of these features across all considered system sizes establishes that the re-entrant localization–delocalization behavior is robust and not a finite-size effect.

To further corroborate these findings, we analyze the fractal dimension $D_2$ under identical conditions, as shown in Fig.~\ref{fig:d33}. For $s=2$, Figs.~\ref{fig:d33}(a) and \ref{fig:d33}(b) display the variation of $D_2$ for the narrow and wide energy windows, respectively, both revealing a clear nonmonotonic dependence on $\lambda$. The corresponding results for $s=3$ are presented in Figs.~\ref{fig:d33}(c) and \ref{fig:d33}(d), where the oscillatory features become more pronounced, indicating an enhanced sensitivity of multifractal characteristics to disorder. Across all system sizes, $D_2$ consistently exhibits alternating regimes of higher and lower values, reflecting repeated transitions between extended and localized states. The robustness of this behavior further confirms that the re-entrant localization–delocalization transitions are an intrinsic property of the system.

\begin{figure}[t]
    \centering
   \includegraphics[width=1.0\linewidth]{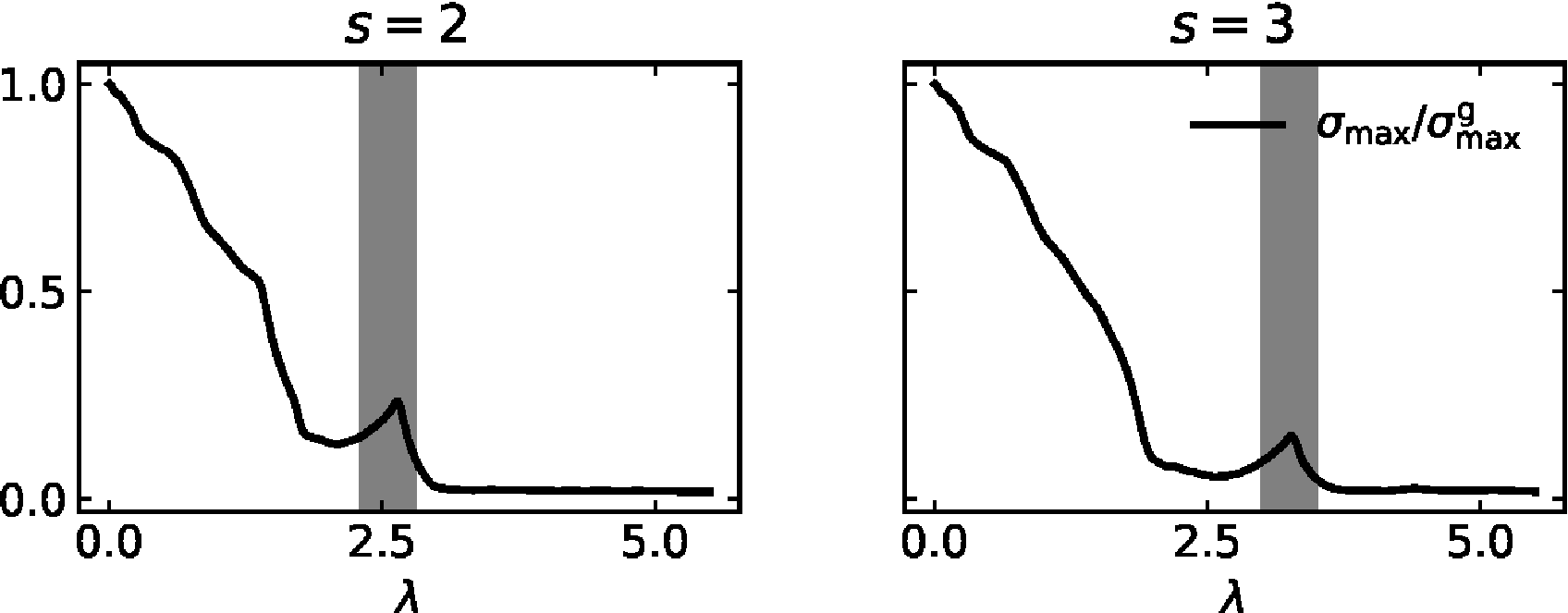}
    \caption{The maximum root-mean-square displacement, $\sigma_{\mathrm{max}}$, normalized by its global maximum, $\sigma_{\mathrm{max}}^{g}$, is plotted as a function of $\lambda$ for $s=2$ and $s=3$, respectively. Here the time evolution is carried out up to $t=200$}
    \label{fig:dyna1}
\end{figure}
\subsection{Dynamical Evidence of Re-entrant Transition}
The time-dependent root-mean-square displacement is defined as
\begin{equation}
\sigma(t) = \left[ \sum_{n} \left(n - \frac{L}{2}\right)^{2} |\psi_{n}(t)|^{2} \right]^{1/2} ,
\end{equation}
where the initial state is localized at the central site $j_0 = L/2$. The maximum extent of spreading during the evolution is quantified by
\begin{equation}
\sigma_{\mathrm{max}} = \max_{t} \, \sigma(t) .
\end{equation}
For a meaningful comparison across parameters, we consider the normalized quantity $\sigma_{\mathrm{max}} / \sigma_{\mathrm{max}}^{g}$ and examine its variation with $\lambda$ for $s=2$ and $s=3$.

From the plot, we observe that the overall behavior closely mirrors that of the average NPR calculated over all eigenstates. In particular, the shaded region highlights a nonmonotonic trend where the quantity, after initially decreasing, exhibits a subsequent increase. This revival is consistent with the features seen in the NPR analysis and provides further evidence of the re-entrant behavior. Notably, the emergence of this trend in the time-evolution dynamics reinforces our earlier conclusions, indicating that the re-entrant localization–delocalization characteristics persist beyond the static eigenstate description and are also manifested dynamically.

\section{Conclusion}

In this work, we have investigated localization in a quasiperiodic diamond lattice with strand-dependent Aubry--André--Harper (AAH) onsite modulations, emphasizing the roles of the modulation ratio $s$ and the averaged potential on the middle strand. By tuning the quasiperiodic strength $\lambda$, we demonstrate the emergence of a robust re-entrant localization behavior, where eigenstates repeatedly alternate between extended and localized phases. This re-entrant transition persists only within a finite range of $s$ and is crucially stabilized by the correlated averaged potential, highlighting their combined role as the key mechanism. 

Our analysis, based on IPR, NPR, and $D_2$, along with the evolution of extended states, finite-size scaling of $\langle \mathrm{NPR} \rangle$ and $\langle D_2 \rangle$, and dynamical signatures from the time-dependent root-mean-square displacement, consistently confirms the robustness of these transitions. The coexistence of extended and localized states and the sensitivity to energy windows further underscore the unconventional nature of the phenomenon. Notably, in contrast to conventional scenarios, the observed re-entrant localization arises solely from strand-dependent onsite quasiperiodic potentials without any hopping modulation. These findings reveal rich localization physics driven by correlated quasiperiodicity and lattice geometry, providing a promising platform for exploring anomalous transport and critical states in engineered systems such as photonic lattices, optical waveguides, and cold-atom setups.

\section*{APPENDIX}
\subsection{Justification of our model}
\begin{figure*}[t]
    \centering
   \includegraphics[width=0.9\linewidth]{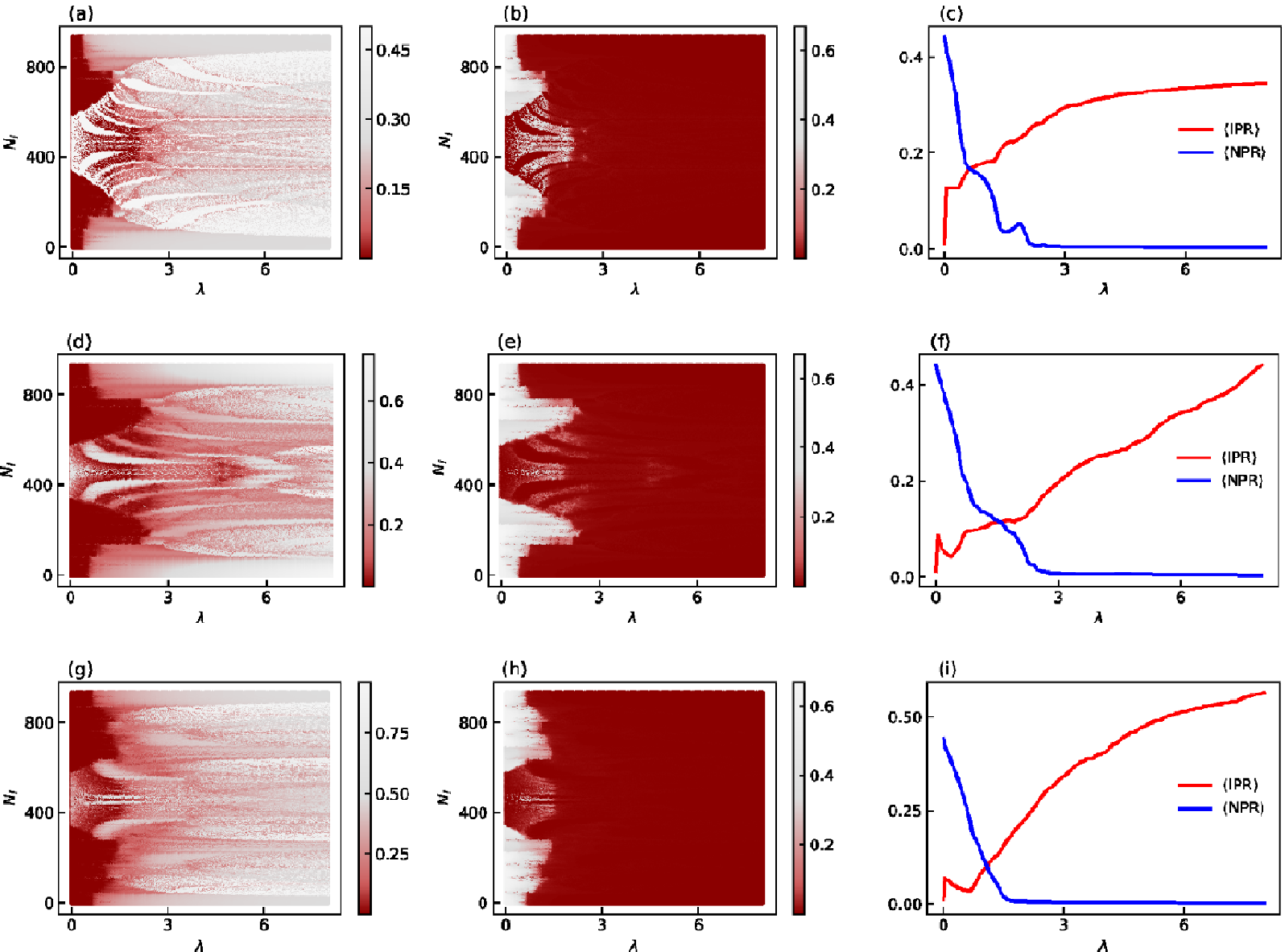}
    \caption{Same layout as Fig.~\ref{fig:state1}. The first and second rows correspond to $s=1$ and $s=6$, respectively. In the third row, keeping $s=3$, the middle strand follows the standard AAH modulation rather than the average of the upper and lower strands. Additionally, within each block the two sites of the middle strand have different on-site energies, unlike in the previous cases.}
    \label{fig:jst1}
\end{figure*}

\begin{figure*}[t]
    \centering
   \includegraphics[width=0.8\linewidth]{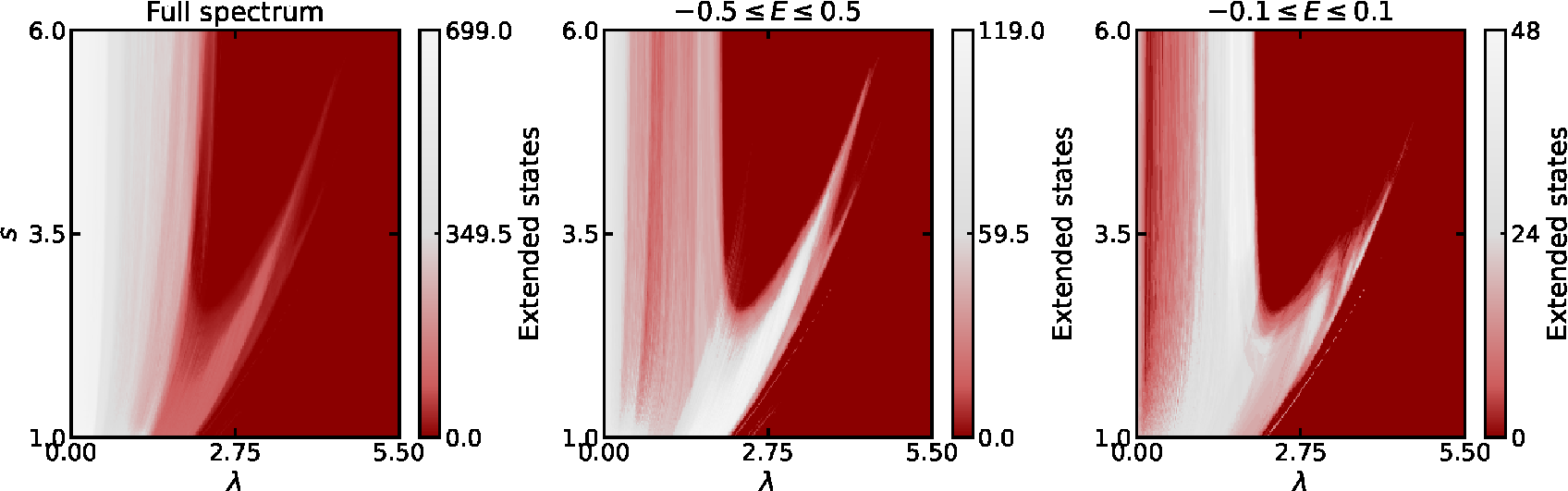}
    \caption{Counting of extended states under the complete parameter space of $s$ and $\lambda$.}
    \label{fig:count2}
\end{figure*}
We next investigate the influence of the parameter $s$ on the re-entrant localization behavior by examining representative cases across its range. For $s=1$ [Figs.~\ref{fig:jst1}(a)–(c)], the eigenstate spectrum, resolved by IPR and NPR, reveals that in the weak-disorder regime the band-center states are predominantly localized, while the band-edge states remain extended. With increasing $\lambda$, this trend gradually reverses, leading to a redistribution of conducting and insulating regions across the spectrum. Notably, a weak signature of re-entrant behavior emerges around $\lambda \approx 2.7$–$3$, where $\langle \mathrm{NPR} \rangle$ exhibits a modest enhancement. However, the absence of a fully suppressed NPR prior to this feature indicates a mixed phase, where localized and extended states coexist, rather than a sharply defined transition. In contrast, for larger values such as $s=6$ [Figs.~\ref{fig:jst1}(d)–(f)], the re-entrant behavior is entirely suppressed. The eigenstate spectrum no longer displays alternating localized and extended regions with increasing $\lambda$, and the averaged quantities $\langle \mathrm{IPR} \rangle$ and $\langle \mathrm{NPR} \rangle$ evolve smoothly without pronounced oscillations, indicating the absence of re-entrant transitions. 

To further elucidate the underlying mechanism, we consider a modified configuration at $s=3$, where the onsite potential on the middle strand follows a standard AAH form instead of the averaged modulation [Figs.~\ref{fig:jst1}(g)–(i)]. In this case, the re-entrant behavior disappears, and the spectral as well as averaged measures no longer exhibit alternating localization–delocalization features. This comparison highlights the essential role of the spatially structured modulation in generating the re-entrant phenomenon. Taken together, these results demonstrate that re-entrant localization is not a generic feature but emerges only within a finite parameter window, approximately $2 \lesssim s \lesssim 5$, where the interplay between quasiperiodicity and lattice geometry is optimally balanced. Beyond this regime, either weak modulation ($s \sim 1$) or excessive modulation ($s \gtrsim 6$) suppresses the re-entrant response, establishing $s$ as a key control parameter governing the emergence and stability of the phenomenon.
\subsection{Variation of extended eigenstate count with $s$ and $\lambda$}
Figure~\ref{fig:count2} presents the number of extended eigenstates as a function of the parameters $s$ and $\lambda$, evaluated over the full spectrum as well as within selected energy windows (first, second, and third columns, respectively). An eigenstate is classified as extended using the criterion $\mathrm{IPR} \leq 0.01$. In the weak-disorder regime, the majority of eigenstates remain extended across all values of $s$, reflecting the dominance of delocalized behavior. As $\lambda$ increases, a systematic reduction in the number of extended states is observed, particularly for smaller values of $s \sim 1$, where the count eventually vanishes, indicating complete localization. However, with increasing $s$, a striking nonmonotonic trend emerges: the number of extended states initially decreases with $\lambda$, followed by a reappearance at intermediate disorder strengths. This behavior provides a clear signature of re-entrant delocalization in the system. The effect is most pronounced within the energy window $-0.5 \leq E \leq 0.5$, although similar tendencies are visible in other spectral regions. Notably, the resurgence of extended states persists only up to a finite range of $s$ (approximately $s \lesssim 6$), beyond which localization dominates. These results demonstrate that the interplay between quasiperiodic modulation and disorder not only induces re-entrant transitions but also allows their extent to be systematically tuned via the parameter $s$.

\subsection{Experimental realization}
A possible experimental realization of the proposed quasiperiodic diamond lattice can be achieved in controllable platforms such as ultracold atoms in optical lattices or photonic waveguide arrays, where lattice geometry and on-site potentials can be engineered with high precision. In an optical lattice setup, interfering laser beams can create a quasi-one-dimensional diamond geometry consisting of three parallel strands connected through diagonal tunneling links, with each unit cell containing four sites: one on the upper strand, one on the lower strand, and two sites forming the middle strand. The quasiperiodic Aubry–André–Harper (AAH) modulation can be introduced by superimposing an additional incommensurate lattice potential along the chain direction. By independently tuning the laser intensities on different strands, the upper strand can experience a quasiperiodic modulation of strength $\lambda$, while the lower strand is subjected to a weaker modulation of amplitude $\lambda/s$. The two sites in the middle strand within each unit cell are designed to have identical on-site potentials, but these potentials vary from block to block following a quasiperiodic profile determined as the average of the potentials applied to the upper and lower strands. Such site-selective control can be realized using spatial light modulators or digital micromirror devices that enable programmable shaping of the lattice potential, making the proposed quasiperiodic diamond lattice experimentally feasible.

\end{document}